\documentclass[reprint,superscriptaddress, amsmath,amssymb, aps, prresearch, longbibliography, floatfix,nofootinbib]{revtex4-2}
\usepackage{amsmath}
\usepackage{graphicx}
\usepackage[normalem]{ulem}
\usepackage{bm}
\usepackage{natbib}
\usepackage{dsfont}
\usepackage{tikz}
\usepackage{epsfig}
\usepackage{feynmf}
\usepackage{blindtext, rotating}
\usepackage{mathtools}
\usepackage{dsfont}
\usepackage{subcaption}
\usepackage{physics}
\usepackage{amsfonts}
\usepackage{xcolor}
\usepackage{ragged2e}
\usepackage{siunitx}
\usepackage{comment}
\usepackage{soul}
\usepackage{empheq}
\usepackage{lipsum}
\usepackage{hyperref} 
\hypersetup{breaklinks=true, colorlinks=true, citecolor=blue, linkcolor=cyan, urlcolor=blue,filecolor=blue}
\usepackage{orcidlink}

\usepackage[T1]{fontenc}

\usepackage{xcolor} 

\DeclareCaptionJustification{justified}{\justifying}

\DeclareSIUnit{\rad}{rad}

\definecolor{bright_blue}{HTML}{85C1E9}
\definecolor{middle_blue}{HTML}{2E86C1}
\definecolor{dark_blue}{HTML}{1B4F72}

\usepackage{hyperref}

\usepackage{orcidlink}

\begin{document}
\title{Lorentz violation and Hawking--Page metastability in Bumblebee--AdS gravity}

\author{Syed Masood\,\texorpdfstring{\orcidlink{0000-0003-3054-7317}}{}}
\email[Work: ]{masood@westlake.edu.cn}
\email[Personal: ]{quantummind137@gmail.com}
\thanks{Author full legal name (official records): Syed Masood Ahmad Shah Bukhari}
\affiliation{Zhejiang University/University of Illinois at Urbana-Champaign (ZJU-UIUC) Institute,
Zhejiang University, International Campus, Haining, Zhejiang 314400, China, \\}
\affiliation{Department of Physics, School of Science and Research Center for Industries of the Future, Westlake University, Hangzhou 310030, P. R. China,\\}
\affiliation{Institute of Natural Sciences, Westlake Institute for Advanced Study, Hangzhou 310024, P. R. China.\\}
\affiliation{Canadian Quantum Research Center, 204-3002 32 Ave, Vernon, BC V1T 2L7, Canada.\\}

\begin{abstract}
The standard Hawking--Page description of AdS black holes, based on an on-shell Gibbs free-energy crossing, does not resolve the metastable black-hole branch or the barrier-mediated conversion between competing phases. We address this issue for neutral AdS black holes in Bumblebee gravity by constructing the generalized off-shell Gibbs free-energy landscape in the canonical ensemble at fixed effective pressure. We find analytically that the minimum temperature \(T_{\min}\) is independent of the Lorentz-violating parameter \(\alpha\), whereas the Hawking--Page temperature \(T_{\mathrm{HP}}\) increases monotonically with \(\alpha\). Consequently, the locally stable but globally disfavored large-black-hole window \(T_{\min}<T<T_{\mathrm{HP}}\) broadens with increasing \(\alpha\). Within the off-shell framework, we derive closed-form expressions for the global-preference gap \(\Delta G_{\mathrm{glob}}\), which measures the free-energy difference between the large-black-hole minimum and thermal AdS, and the barrier height \(\Delta G_{\mathrm{bar}}\), which measures the separation between the small-black-hole saddle and the large-black-hole minimum. We recover the Schwarzschild--AdS limit as \(\alpha\to0\). These results establish the off-shell free-energy landscape as a precise thermodynamic framework for diagnosing Lorentz-violating effects in AdS black-hole phase structure.
\end{abstract}

\date{\today}
\maketitle

\section{Introduction}\label{intro}

Lorentz symmetry is one of the most precisely tested principles of modern physics~\cite{Mariz:2022oib,Tasson:2014dfa}. Nevertheless, a broad class of ultraviolet completions of gravity—including string theory~\cite{Kostelecky:1988zi}, loop quantum gravity~\cite{Alfaro:2001rb,Gambini:1998it}, noncommutative field theory~\cite{Carroll:2001ws}, and Ho\v{r}ava--Lifshitz gravity~\cite{Horava:2009uw}—predict low-energy remnants of Lorentz violation (LV), typically near the Planck scale~\cite{Kostelecky:2003fs}. The Standard Model Extension (SME)~\cite{Liberati:2013xla} provides a systematic effective-field-theory framework in which such deformations can be parametrized and constrained observationally~\cite{Tasson:2014dfa,Addazi:2021xuf,LHAASO:2024lub}.

In the gravitational sector, spontaneous LV arises when a dynamical tensor field acquires a nonzero vacuum expectation value (VEV), thereby selecting a preferred spacetime direction. The vector case—commonly referred to as Bumblebee gravity~\cite{Kostelecky:1989jw,Kostelecky:1989jp,Bailey:2006fd,Bluhm:2008yt}—has become a controlled theoretical laboratory for studying the phenomenological consequences of LV in curved spacetime. Spherically symmetric black hole solutions in Bumblebee gravity were first derived in Ref.~\cite{Casana:2017jkc} and subsequently extended to include a cosmological constant in Ref.~\cite{Maluf:2020kgf}, yielding a one-parameter family of solutions in which $\alpha$—the LV coupling—enters through an effective rescaling of the AdS curvature scale. Various observational aspects of this solution, including thermodynamics~\cite{Mai:2023ggs,An:2024fzf}, Hawking radiation~\cite{Kanzi:2019gtu}, quasinormal modes~\cite{Zhang:2023wwk}, gravitational lensing~\cite{Ovgun:2018ran}, and Event Horizon Telescope constraints~\cite{Xu:2023xqh}, have been studied.\\
\indent Despite this progress, the thermodynamic phase structure of Bumblebee--AdS black holes—and specifically how LV modifies the organization of competing thermodynamic phases—has not been analyzed with the resolution provided by the off-shell free-energy landscape. In the canonical ensemble, the Hawking--Page (HP) transition~\cite{Hawking:1982dh} marks the coexistence boundary between thermal AdS radiation and a large stable black hole. The on-shell analysis, based on the sign of the Gibbs free energy, identifies the transition temperature but does not resolve the metastable coexistence region $T_{\min}<T<T_{\rm HP}$ where a locally stable large black hole (LBH) and globally preferred thermal AdS coexist. The off-shell free-energy landscape~\cite{Spallucci:2013jja,Li:2020khm}—constructed by promoting the horizon radius to an order parameter and allowing off-shell fluctuations—makes this structure explicit through the global-preference gap $\Delta G_{\rm glob}$ and the free-energy barrier $\Delta G_{\rm bar}$~\cite{Ali:2023wkq,Li:2023men,Liu:2023sbf}. The free energy of the non-equilibrium black hole configurations in this landscape has been given a concrete path-integral foundation via Euclidean gravitational instantons with conical singularities~\cite{Li:2022oup}.\\
\indent In this work, we use the Bumblebee--AdS black hole~\cite{Maluf:2020kgf} as a minimal, analytically tractable model to probe the interplay between LV deformations and thermodynamic phase structure. Our results indicate that the HP temperature $T_{\rm HP}$ shifts monotonically with $\alpha$, while $T_{\min}$ remains exactly independent of $\alpha$, structurally widening the metastable window.  The off-shell landscape quantifies the resulting LV-deformed barrier and global-preference gap analytically. Finally, the standard Schwarzschild--AdS phase structure is recovered smoothly as $\alpha\to 0$.

The paper is organized as follows. Section~\ref{geom} introduces the Bumblebee--AdS geometry and its horizon structure and develops the on-shell thermodynamics, including heat capacity and Gibbs free energy. Section~\ref{sec:HPT} identifies the $\alpha$-shifted Hawking--Page benchmark. Section~\ref{sec:GFLV} presents the off-shell landscape analysis, including the overlap criterion, global-preference gap, and barrier height. Section~\ref{summary} summarizes our findings.

\section{Model and on-shell thermodynamics}\label{geom}
\subsection{Geometric structure}
Throughout we set $c=G=\hbar=1$.
The static, spherically symmetric line element is
\begin{equation}
ds^{2}=g_{tt}(r)\,dt^{2}+g_{rr}(r)\,dr^{2}+r^{2}d\Omega_{2}^{2},
\end{equation}
with the metric function~\cite{Maluf:2020kgf}
\begin{equation}\label{metric}
g_{tt}(r)=-f(r),\qquad
f(r)=1-\frac{2M}{r}-\frac{(1+\alpha)\Lambda_{e}}{3}\,r^{2},
\end{equation}
where $\alpha$ is the LV parameter, $\Lambda_{e}=\Lambda/(1+\alpha)$ acts as an effective cosmological constant.
The event horizon radius is the largest positive root of $f(r_+)=0$, i.e.
\begin{equation}\label{eq:horizon_cubic}
\frac{(1+\alpha)\Lambda_{e}}{3}\,r_{+}^{3}-r_{+}+2M=0,
\end{equation}
for which a closed-form Cardano expression exists (see Appendix \ref{APPENDA}).

Figure~\ref{rgplot} displays $r_+(\Lambda_e,\alpha)$ at fixed $M$.
Two monotonicities follow directly from \eqref{eq:horizon_cubic} by implicit differentiation:
\begin{align}
\frac{\partial r_{+}}{\partial \alpha}
&=-\frac{\partial_{\alpha}f}{\partial_{r}f}\bigg|_{r=r_+}
=-\frac{-\Lambda_{e}r_{+}^{2}/3}{2M/r_{+}^{2}-2(1+\alpha)\Lambda_{e}r_{+}/3},
\\
\frac{\partial r_{+}}{\partial \Lambda_{e}}
&=-\frac{\partial_{\Lambda_e}f}{\partial_{r}f}\bigg|_{r=r_+}
=-\frac{-(1+\alpha)r_{+}^{2}/3}{2M/r_{+}^{2}-2(1+\alpha)\Lambda_{e}r_{+}/3}.
\end{align}
For AdS ($\Lambda_e<0$) the denominator is strictly positive at the outer horizon, hence
\begin{equation}
\frac{\partial r_+}{\partial \alpha}<0,\qquad
\frac{\partial r_+}{\partial \Lambda_e}>0,
\end{equation}
explaining the observed decrease of $r_+$ with increasing $\alpha$ at fixed $\Lambda_e$ and the
increase of $r_+$ as $\Lambda_e\to 0^{-}$ in Fig.~\ref{rgplot}.
In particular, the Schwarzschild limit is recovered smoothly:
\begin{equation}
r_+=2M+\frac{8}{3}(1+\alpha)\Lambda_e M^{3}+O(\Lambda_e^{2}),
\end{equation}
so $r_+\to 2M$ and the $\alpha$-dependence disappears as $\Lambda_e\to 0$,
consistent with the nearly horizontal band at the top of Fig.~\ref{rgplot}.

\begin{figure}[t]
  \centering
\includegraphics[width=\columnwidth]{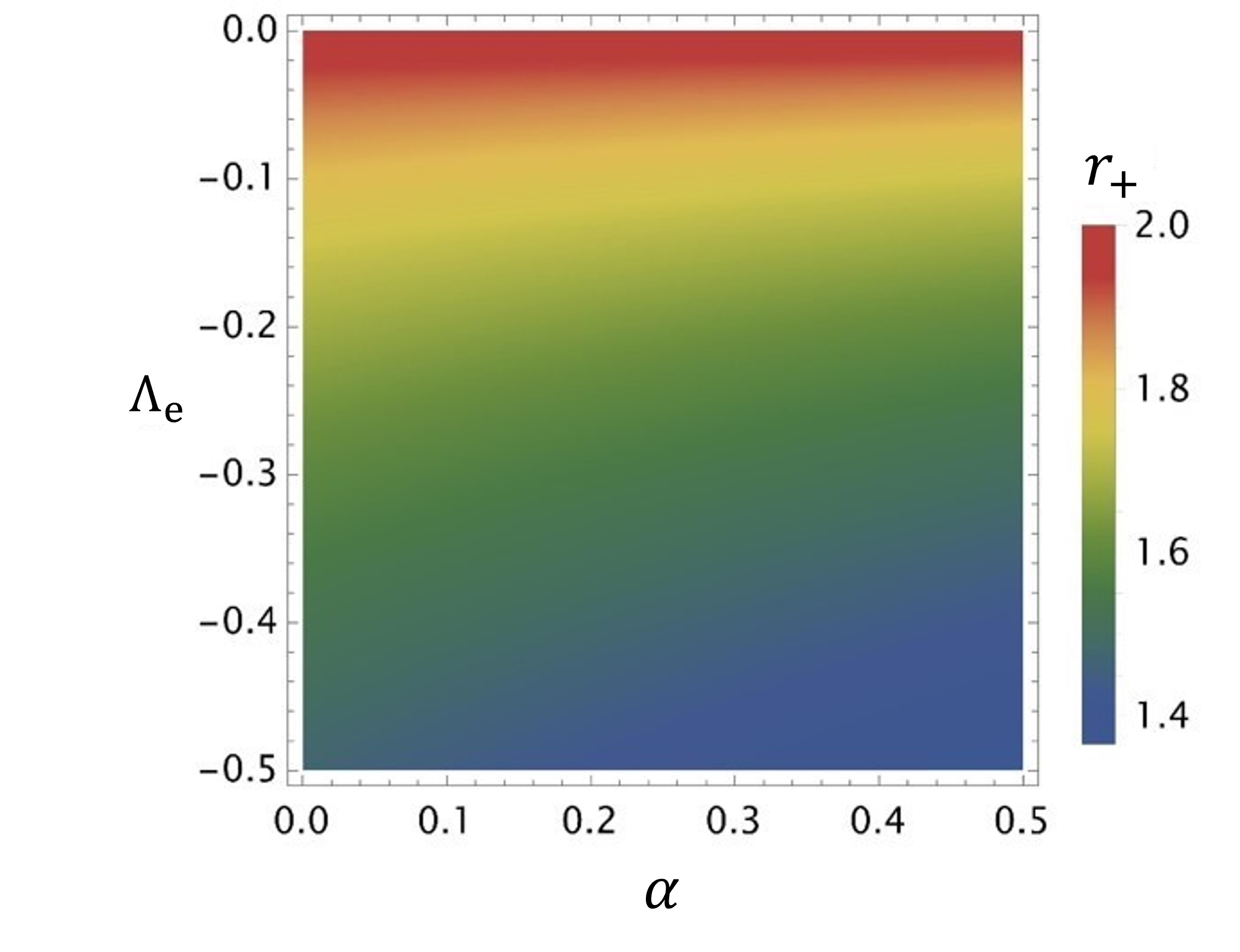}
  \caption{\label{rgplot} Horizon structure of black holes in Bumblebee gravity, with LV effects via $\alpha$. We set $M=1$.}
\end{figure}

The parameter $\alpha$ does not introduce an additional radial term in the metric function $f(r)$; instead, it modifies the AdS contribution through the single combination $(1+\alpha)\Lambda_e$ appearing in the horizon equation. Consequently, at fixed $(M,\Lambda_e)$, increasing $\alpha$ shifts the outer horizon to smaller radius $r_+$.

We restrict to $\alpha>0$ in this work; negative $\alpha$ can be treated analogously.

\begin{figure}[b]
\centering
\includegraphics[width=\columnwidth]{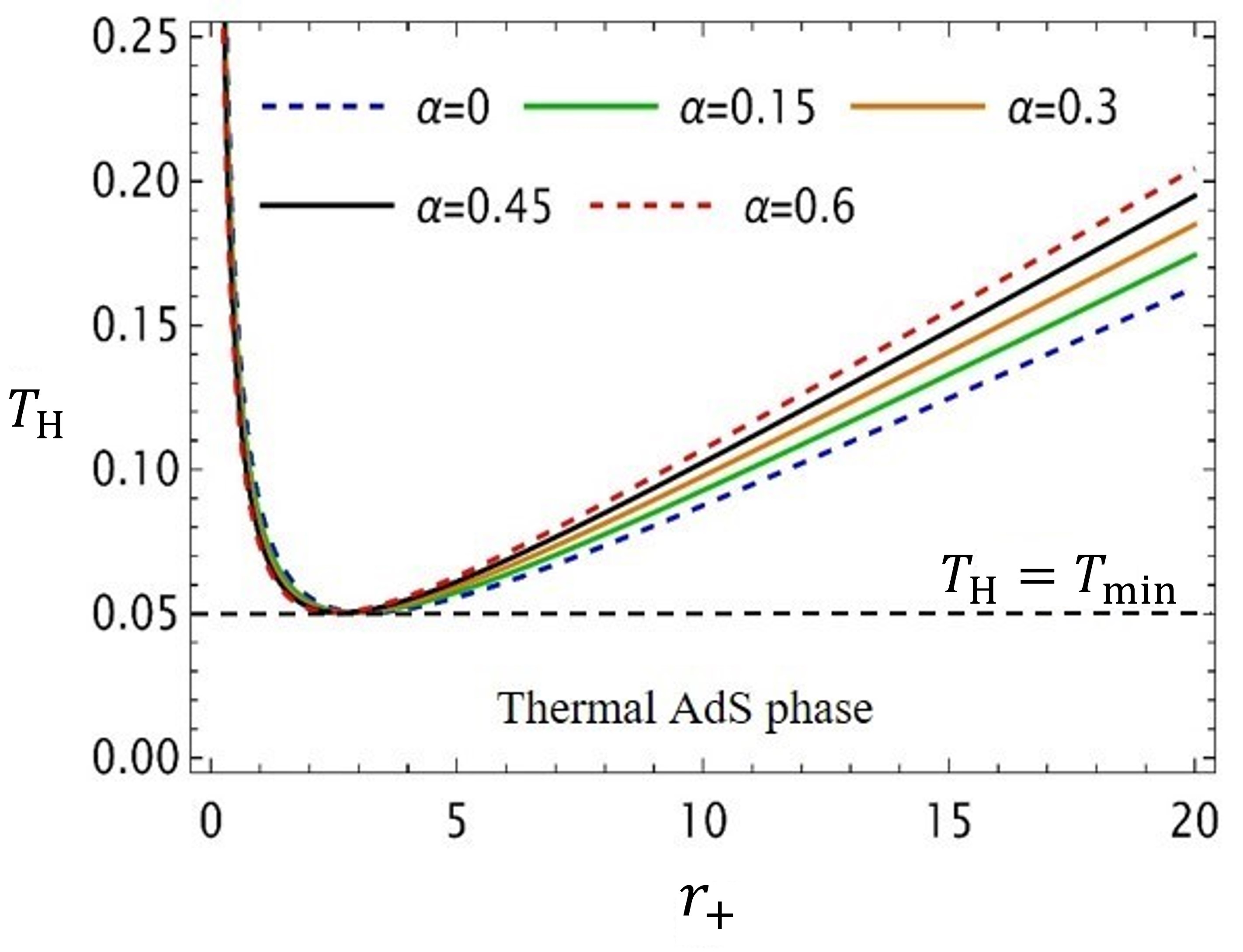}
\caption{\label{Tplot} Hawking temperature $T_{\rm H}$ vs. $r_{+}$ for various $\alpha$. We set $\Lambda_{e}=-0.1$ across all cases.}%
\end{figure}

\subsection{Black hole branching}

The Hawking temperature follows from the surface gravity associated with the timelike Killing field
$\eta^\mu=(\partial_t)^\mu$,
\begin{align}
T_{\rm H}=\frac{\kappa}{2\pi}
=-\frac{1}{4\pi\sqrt{-g_{tt}g_{rr}}}\,\frac{\partial g_{tt}}{\partial r}\bigg|_{r=r_+}.
\label{temp}
\end{align}
From Eq.~(\ref{metric}) one obtains
\begin{align}
T_{\rm H}(r_+)=\frac{1-(1+\alpha)\Lambda_e\,r_+^2}{4\pi\sqrt{1+\alpha}\,r_+}.
\label{TEMP1}
\end{align}
For the AdS branch $\Lambda_e<0$, it is useful to rewrite \eqref{TEMP1} as
\begin{align}
T_{\rm H}(r_+)=\frac{1}{4\pi\sqrt{1+\alpha}}
\left(\frac{1}{r_+}+(1+\alpha)|\Lambda_e|\,r_+\right),
\label{TEMP1_AdS_form}
\end{align}
from which one can see that for  $r_+\to 0$, $T_{\rm H}\sim [4\pi\sqrt{1+\alpha}\,r_+]^{-1}\to\infty$, while for $r_+\to\infty$, one finds
\begin{align}
T_{\rm H}(r_+)=\frac{\sqrt{1+\alpha}\,|\Lambda_e|}{4\pi}\,r_+
+\frac{1}{4\pi\sqrt{1+\alpha}}\frac{1}{r_+}+\mathcal{O}(r_+^{-3}),
\label{T_asymp}
\end{align}
so the large-$r_+$ branch grows linearly with asymptotic slope
$\mathrm{d}T_{\rm H}/\mathrm{d}r_+\to \sqrt{1+\alpha}\,|\Lambda_e|/(4\pi)$, which increases monotonically with $\alpha$.
This accounts for the observed fan-out of the curves in Fig.~\ref{Tplot} at large $r_+$: larger $\alpha$ corresponds to a steeper large-black-hole branch.

Since $T_{\rm H}\to\infty$ both as $r_+\to0$ and as $r_+\to\infty$, $T_{\rm H}(r_+)$ admits a unique global minimum,
determined by $\mathrm{d}T_{\rm H}/\mathrm{d}r_+=0$:
\begin{align}
r_{\min}=\frac{1}{\sqrt{-(1+\alpha)\Lambda_e}}.
\label{rmin_caseI}
\end{align}
corresponding to the temperature
\begin{align}
T_{\min}=T_{\rm H}(r_{\min})
=\frac{\sqrt{-\Lambda_e}}{2\pi}.
\label{TmincaseI}
\end{align}
The radius (\ref{rmin_caseI}) defines the standard small-/large-black-hole branching:
we refer to the branch $r_+<r_{\min}$ as the small black hole (SBH) branch and $r_+>r_{\min}$ as the large black hole (LBH) branch.
In the canonical ensemble, black-hole solutions exist only for $T\ge T_{\min}\equiv T_{\rm H}(r_{\min})$, while for $T<T_{\min}$ the ensemble is described by thermal AdS radiation \cite{Hawking:1982dh}. Below we will see that this same point $r_{\rm min}$ coincides with the divergence of the heat capacity, marking the change of local thermal stability between the two branches.

The $\alpha$-independence of $T_{\rm min}$ in (\ref{TmincaseI}) follows from a simple cancellation.
At the stationary point the two contributions in \eqref{TEMP1_AdS_form} are equal, such that
\begin{align}
\frac{1}{r_{\min}}=(1+\alpha)|\Lambda_e|\,r_{\min},
\end{align}
and using $r_{\min}=1/\sqrt{(1+\alpha)|\Lambda_e|}$ we also have
\begin{align}
\frac{1}{r_{\min}}=\sqrt{(1+\alpha)|\Lambda_e|}.
\end{align}
so the parentheses in \eqref{TEMP1_AdS_form} evaluate to $2\sqrt{(1+\alpha)|\Lambda_e|}$.
The prefactor $1/\sqrt{1+\alpha}$ removes the $\alpha$-dependence in (\ref{TmincaseI}). Accordingly, increasing $\alpha$ shifts the branching point to smaller radius, $r_{\min}\propto (1+\alpha)^{-1/2}$, without changing $T_{\min}$,
while it steepens the LBH branch through the asymptotic slope \eqref{T_asymp}.

\subsection{Stability criteria: local response vs global preference}
We work in the canonical ensemble at fixed effective pressure
$P\equiv -\Lambda_{e}/(8\pi)$ and use $r_{+}$ as the order parameter.
Local thermal stability is governed by the sign of a response function, while
global phase preference is fixed by the on-shell Gibbs free energy relative to
thermal AdS, $G_{\rm AdS}=0$ \cite{Hawking:1982dh,Akbar:2004ke}. 

\paragraph{Local response: heat capacity.}
\begin{figure}[t]
\includegraphics[width=\columnwidth]{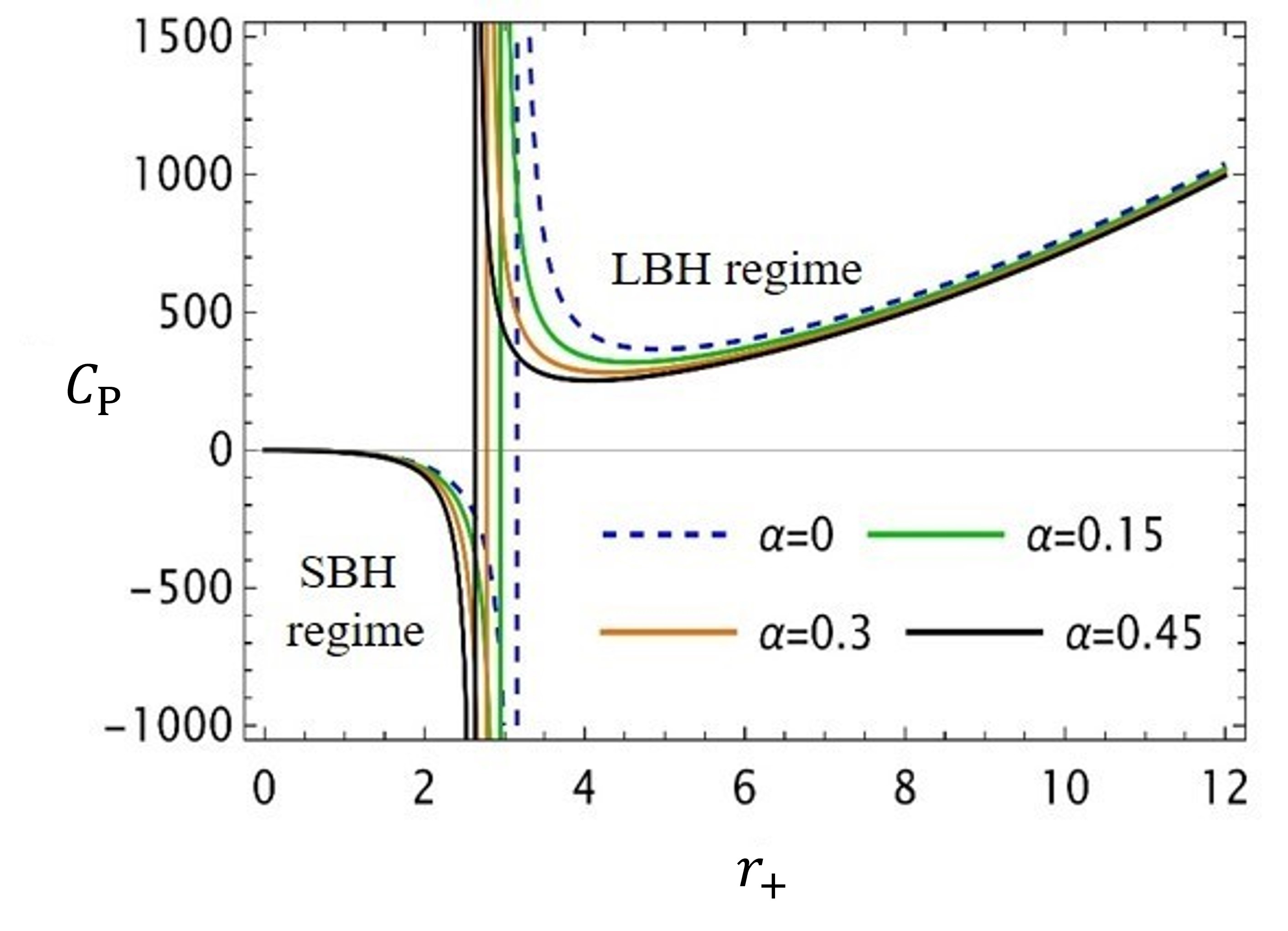}
\caption{\label{Cplot} Heat capacity of the black hole with various choices of $\alpha$: The vertical lines depict the point $r_{\mathrm{min}}$ for all the cases. Note that by fixing $\Lambda_{e}$, one has to treat above plots as effective description. We chose $\Lambda_{e}=-1$. }
\end{figure}
\begin{figure}
\includegraphics[width=\columnwidth]{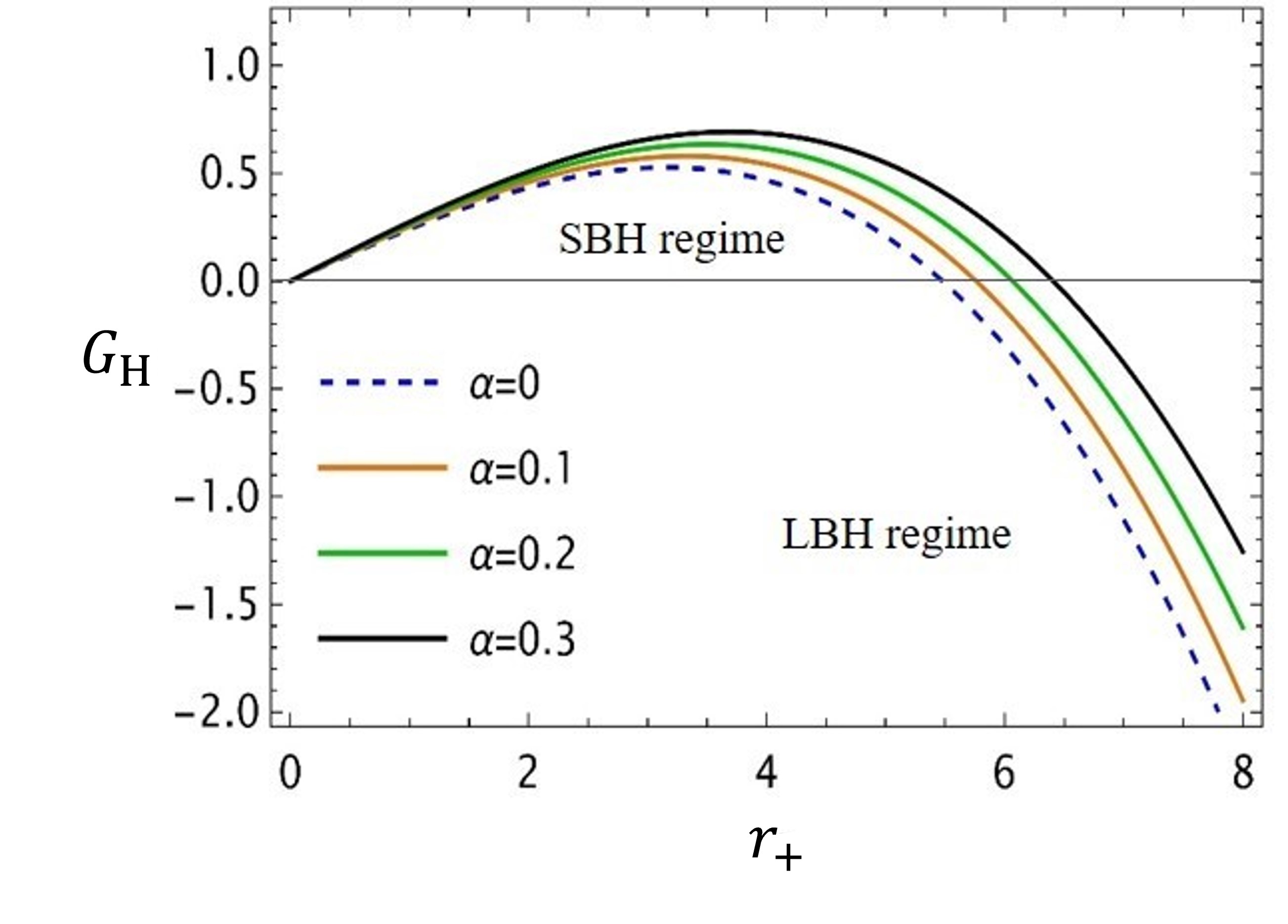}
\caption{\label{Fplot} On-shell free energy $G_{\rm H}$ vs black hole size $r_{+}$ We chose $\Lambda_{e}=-1$. }
\end{figure}
At fixed $P$ the heat capacity can be written in the form
\begin{equation}
C_{\rm P}\equiv T\left(\frac{\partial S}{\partial T}\right)_{P}
= T\,\frac{(\partial S/\partial r_{+})_{P}}{(\partial T/\partial r_{+})_{P}},
\end{equation}
so that divergences of $C_{\rm P}$ coincide with turning points of the equation of
state $T_{\rm H}(r_{+})$ (spinodals) \cite{Davies:1978zz,Akbar:2004ke}.
For our LV--AdS solutions one finds
\begin{equation}\label{Cmaluf}
C_{\rm P}(r_{+})=\frac{2\pi r_{+}^{2}\Big[(1+\alpha)\Lambda_{e}r_{+}^{2}-1\Big]}
{(1+\alpha)\Lambda_{e}r_{+}^{2}+1}\, .
\end{equation}
The pole is located at
\begin{equation}\label{rmin}
(1+\alpha)\Lambda_{e}r_{+}^{2}+1=0
\quad\Longrightarrow\quad
r_{+}=r_{\min}=\frac{1}{\sqrt{-(1+\alpha)\Lambda_{e}}}\, 
\end{equation}
which is precisely the SBH/LBH branching point, with $C_{\rm P}<0$ on the SBH branch
and $C_{\rm P}>0$ on the LBH branch \cite{Davies:1978zz,Akbar:2004ke}.
At fixed $\Lambda_{e}$ the LV dependence enters through the combination
$\Xi\equiv (1+\alpha)\Lambda_{e}$, implying $r_{\min}\propto (1+\alpha)^{-1/2}$.
This explains the leftward shift of the vertical asymptote in Fig.~\ref{Cplot}
as $\alpha$ increases.
Moreover, on the LBH branch ($\Xi r_{+}^{2}<-1$) the factor
$[(\Xi r_{+}^{2}-1)/(\Xi r_{+}^{2}+1)]$ monotonically decreases towards unity as
$\alpha$ grows (at fixed $r_{+}$ and $\Lambda_{e}$), yielding the systematic
suppression of $C_{\rm P}$ seen in Fig.~\ref{Cplot} while preserving the large-$r_{+}$
asymptotic scaling $C_{\rm P}\sim 2\pi r_{+}^{2}$.

\paragraph{Global preference: on-shell Gibbs free energy.}
Global stability at fixed $(T,P)$ is governed by the Gibbs free energy
\begin{equation}
G_{\rm H}\equiv M-T_{\rm H}S,
\end{equation}
evaluated on-shell. The condition $-g_{tt}=0$ yields mass  
\begin{eqnarray}
    M(r_{+})&=\frac{1}{6} r_{+} \left[3-(1+\alpha) \Lambda_{e}  r_{+}^2\right],
\end{eqnarray}
which satisfies extended first law of black hole thermodynamics $\mathrm{d}M=T_{H}\rm{d}S+V\rm{d}P$ and along with entropy $S=\pi r_{+}^2$ and $T_{H}$ yields 
\begin{equation}\label{GH_maluf}
G_{\rm H}(r_{+})
=-\frac{r_{+}\Big[1-(1+\alpha)\Lambda_{e}r_{+}^{2}\Big]}{4\sqrt{1+\alpha}}
-\frac{r_{+}}{6}\Big[(1+\alpha)\Lambda_{e}r_{+}^{2}-3\Big].
\end{equation}
which is plotted in Fig.\ref{Fplot}. $G_{\rm H}$ is positive at small $r_{+}$ and dominated by a negative cubic term at large $r_{+}$, hence it exhibits a single zero
$G_{\rm H}(r_{\rm HP})=0$ which defines the Hawking--Page coexistence point
\cite{Hawking:1982dh,Akbar:2004ke}.
Importantly, in contrast to \eqref{Cmaluf}, the LV dressing of \eqref{GH_maluf}
is \emph{not} exhausted by $\Xi=(1+\alpha)\Lambda_{e}$: the explicit prefactor
$(1+\alpha)^{-1/2}$ produces an additional, inequivalent $\alpha$-dependence in
the linear and cubic contributions.
At fixed $\Lambda_{e}$ this lifts the entire $G_{\rm H}(r_{+})$ curve with increasing
$\alpha$ (Fig.~\ref{Fplot}), 
thereby shifting the root to larger $r_{\rm HP}$ and
enlarging the interval $r_{\min}<r_{+}<r_{\rm HP}$ in which the LBH is locally
stable ($C_{P}>0$) but globally disfavored ($G_{\rm H}>0$); this metastability window
is the regime naturally captured by the off-shell landscape analysis in
Sec.~\ref{sec:GFLV}.

\section{On-shell Hawking–Page transition}\label{sec:HPT}

In the canonical ensemble at fixed effective pressure (equivalently fixed $\Lambda_e<0$), the globally preferred phase at a given temperature is determined by the Gibbs free energy.
We normalize the reference thermal AdS background such that $G_{\rm AdS}=0$ \cite{Hawking:1982dh}, and compare it with the on-shell black-hole Gibbs free energy
\begin{align}
G_{\rm H}(r_+) \,=\, M(r_+) - T_{\rm H}(r_+)\,S(r_+),
\label{GH_def}
\end{align}
evaluated on the black-hole branch.
The Hawking--Page (HP) transition occurs when the two phases have equal free energy,
\begin{align}
G_{\rm H}(r_{\rm HP})=G_{\rm AdS}=0,
\label{HP_condition}
\end{align}
at which the system undergoes a first-order transition between thermal AdS and a large AdS black hole \cite{Hawking:1982dh}.

\begin{figure}[b]
\centering
\includegraphics[width=\columnwidth]{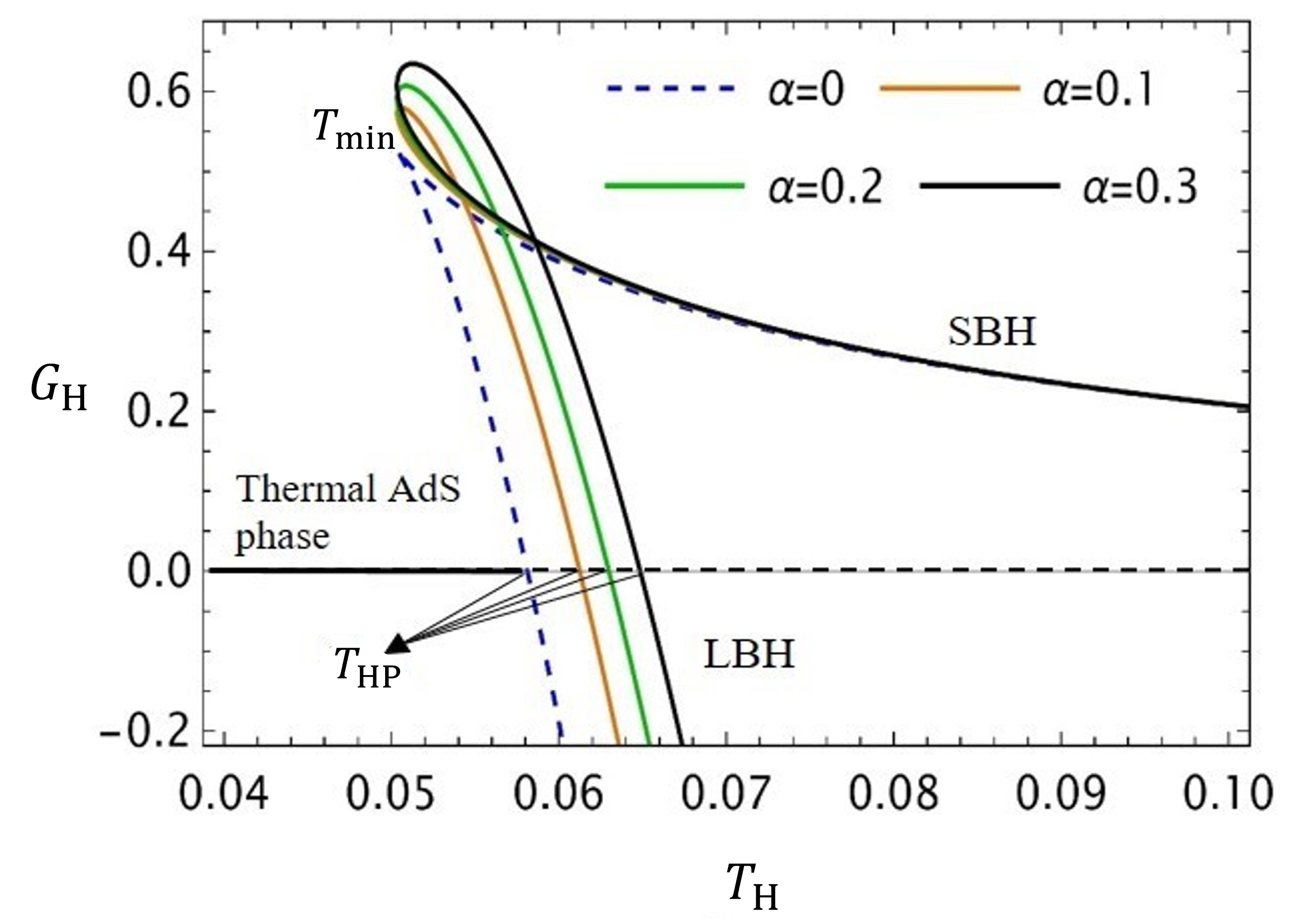}
\caption{Parametric Gibbs free energy $G_{\rm H}$ versus Hawking temperature $T_{\rm H}$ for the Bumblebee--AdS black hole, for several values of the LV parameter $\alpha$ at fixed $\Lambda_e$.
The cusp marks $T_{\min}$ (SBH/LBH branching), while the intersection with $G_{\rm H}=0$ determines the Hawking--Page temperature $T_{\rm HP}$.}
\label{GTplots}
\end{figure}

A convenient graphical diagnosis is provided by the parametric plot of $G_{\rm H}$ versus $T_{\rm H}$ with $r_+$ as the parameter, shown in Fig.~\ref{GTplots} for several values of $\alpha$.
The cusp corresponds to the minimum temperature $T_{\min}$ where ${\rm d}T_{\rm H}/{\rm d}r_+=0$ (equivalently the SBH/LBH branching point discussed in Sec.~\ref{geom}),
so that for $T<T_{\min}$ no on-shell black-hole solution exists and the ensemble is described by thermal AdS radiation.
The minimum temperature is \emph{exactly independent} of the LV parameter $\alpha$: evaluating $T_{\rm H}$ at $r_{\min}$ yields $T_{\min}=\sqrt{-\Lambda_e}/(2\pi)$
Thus, changing $\alpha$ shifts the location of the cusp in $r_+$ (through $r_{\min}\propto (1+\alpha)^{-1/2}$) while leaving its temperature coordinate fixed.
Any small visible offset of the cusp height in Fig.~\ref{GTplots} is therefore numerical/graphical resolution rather than a physical effect.

For $T>T_{\min}$ there are two on-shell branches at fixed $T$: the small-black-hole (SBH) branch with $r_+<r_{\min}$ and the large-black-hole (LBH) branch with $r_+>r_{\min}$.
Local stability is controlled by the heat capacity, with SBH unstable and LBH locally stable, while global preference is determined by the sign of $G_{\rm H}$ relative to $G_{\rm AdS}=0$.\\
\indent For Bumblebee--AdS black holes the HP point follows analytically from \eqref{HP_condition}.
Using the explicit expression for $G_{\rm H}$ from Eq.~(\ref{GH_maluf}), one finds the HP radius
\begin{align}
r_{\rm HP}=\sqrt{-\frac{3}{(1+\alpha)\Lambda_e}}.
\label{rHP}
\end{align}
Substituting \eqref{rHP} into the Hawking temperature \eqref{TEMP1} yields the Hawking--Page temperature
\begin{align}
T_{\rm HP}=T_{\rm H}(r_{\rm HP})
=\frac{\sqrt{(1+\alpha)(-\Lambda_e)}}{\sqrt{3}\,\pi}.
\label{THP}
\end{align}
Equivalently, in extended thermodynamics with $P=-\Lambda_e/(8\pi)$, this can be written as
\begin{eqnarray}\label{THP1}
T_{\rm HP}=\sqrt{1+\alpha}\,\sqrt{\frac{8P}{3\pi}}.
\label{THP_P}
\end{eqnarray}
Thus, at fixed $\Lambda_e$ (or fixed $P$) the HP temperature increases monotonically with $\alpha$, consistent with the rightward shift of the $G_{\rm H}=0$ crossing in Fig.~\ref{GTplots}. It is noteworthy that the interval
\begin{align}
T_{\min}<T<T_{\rm HP}
\label{metastable_window}
\end{align}
is precisely the regime in which an LBH configuration exists and is locally stable, yet thermal AdS remains globally preferred ($G_{\rm H}>0$ on the LBH branch).
In the next section we quantify this metastability using the off-shell free-energy landscape, by computing the global-preference gap and the activation barrier separating the SBH saddle from the LBH minimum.

\begin{figure*}[t]
\centering
\includegraphics[width=\textwidth, height=11cm]{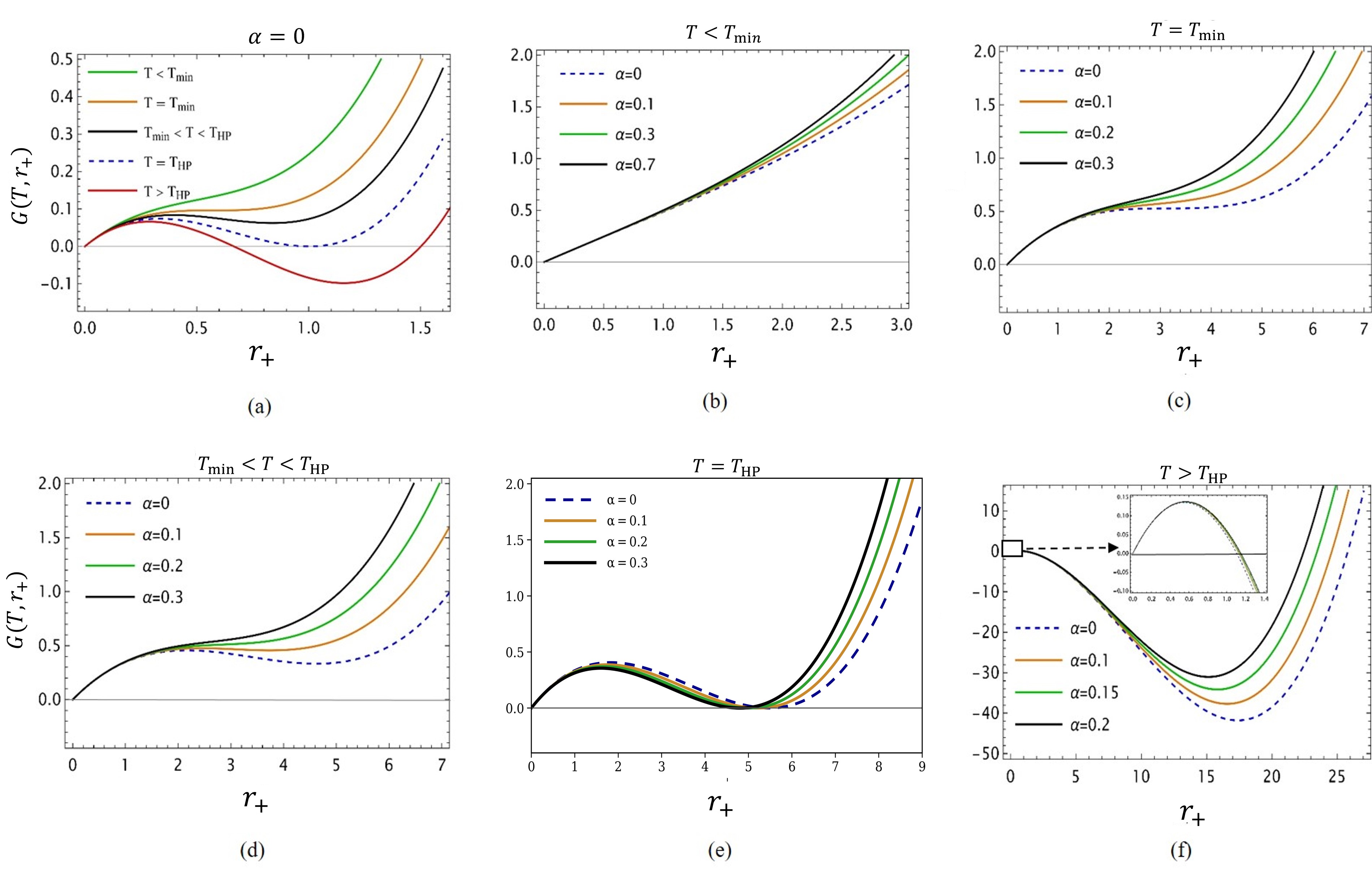}
\caption{ Generalized free energy landscape Bumblebee--AdS gravity. We chose $\Lambda_{e}=-0.1$}%
\label{Gmplot}
\end{figure*}

\begin{figure*}[tb] 
  \centering
  \includegraphics[width=0.8\textwidth,height=5.2cm]{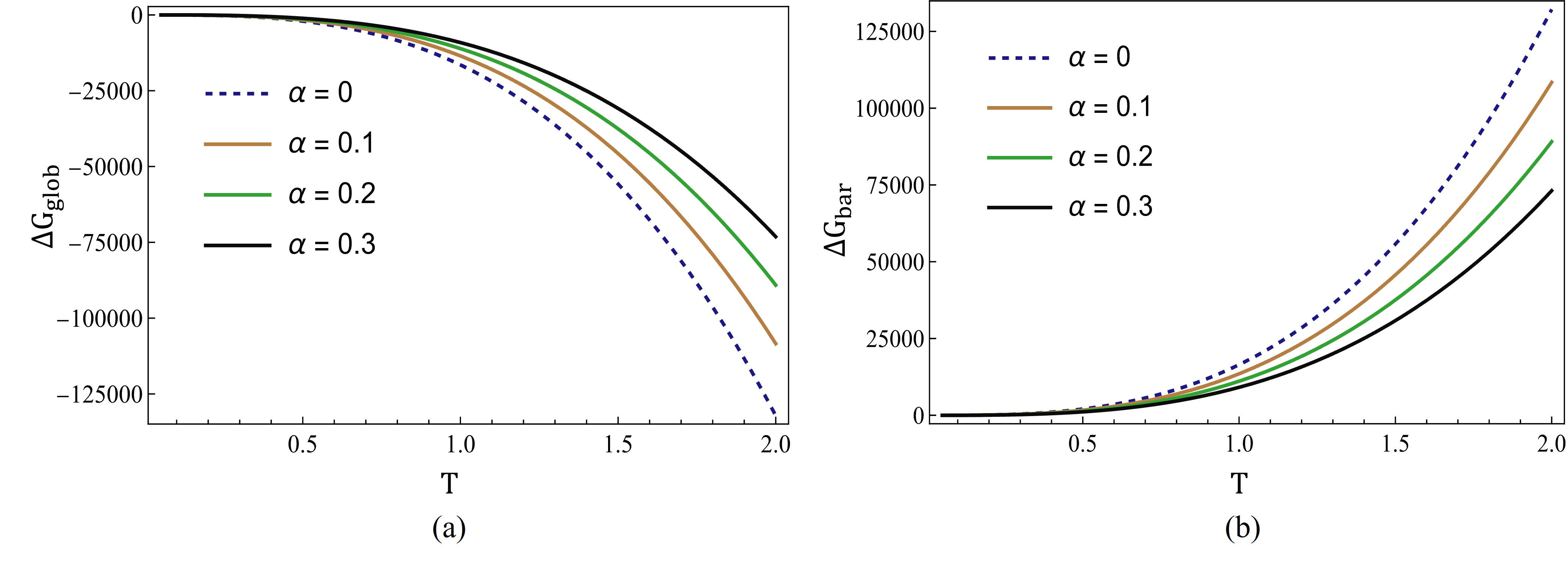}
  \caption{(a) Global preference gap $\Delta G_{\rm glob}$ and  (b) barrier height $\Delta G_{\rm bar}$ for Bumblebee gravity. We set $\Lambda_{\rm e}=-0.1$}
  \label{fig:Deltaglob}
\end{figure*}

\subsection{Stability-scale splitting and metastability width}

A distinctive feature of the Bumblebee--AdS solution is that the temperatures governing black-hole branching and Hawking--Page coexistence respond differently to the LV coupling. From Eqs.~\eqref{TmincaseI} and \eqref{THP1}, one has
\begin{equation}
T_{\min}=\frac{\sqrt{-\Lambda_e}}{2\pi},
\qquad
T_{\rm HP}(\alpha)=\frac{\sqrt{(1+\alpha)(-\Lambda_e)}}{\sqrt{3}\,\pi}.
\end{equation}
This motivates the dimensionless stability-splitting ratio
\begin{equation}
\mathcal{S}(\alpha)\equiv \frac{T_{\rm HP}(\alpha)}{T_{\min}}
=\frac{2}{\sqrt{3}}\sqrt{1+\alpha},
\end{equation}
and the associated normalized metastability width
\begin{equation}
\mathcal{W}(\alpha)\equiv \frac{T_{\rm HP}(\alpha)-T_{\min}}{T_{\min}}
=\frac{2}{\sqrt{3}}\sqrt{1+\alpha}-1.
\end{equation}
The quantity $\mathcal{W}$ measures the separation between the onset of black-hole existence/local stability and the onset of global preference reversal. In Bumblebee gravity this separation is exact, monotonic in $\alpha$, and independent of $\Lambda_e$, showing that Lorentz violation does not merely shift the Hawking--Page point, but splits the local and global thermodynamic stability scales. Table~\ref{tab:HP_compare} summarizes the resulting stability-scale splitting relative to the Schwarzschild--AdS case. The key distinction is that the LV deformation leaves $T_{\min}$ invariant while shifting $T_{\rm HP}$, thereby enlarging the metastable Hawking--Page window in a controlled and exactly quantifiable manner.

\section{Off-shell landscape and overlap criterion}\label{sec:GFLV}

The thermodynamic phase structure discussed above can be analyzed in a canonical ensemble using the generalized (off-shell) free-energy landscape with the free energy given by approach~\cite{Spallucci:2013jja,Li:2020khm}
the off-shell Gibbs free energy reads
\begin{equation}
G(T,r_+)=\frac{1}{6}r_+\Bigl[3-(1+\alpha)\Lambda_e r_+^2\Bigr]-\pi T\, r_+^2,
\label{GI_offshell}
\end{equation}

In this framework, the order parameter is the horizon radius $r_+\ge 0$, with thermal AdS represented by the reference branch at $r_+=0$ and
\begin{equation}
G_{\rm AdS}=0.
\end{equation}
For a fixed ensemble temperature $T$, black-hole configurations are described by the off-shell Gibbs free energy $G(T,r_+)$, while on-shell black-hole phases correspond to stationary points of $G$.

At fixed $(T,\Lambda_e,\alpha)$, stationary points satisfy
\begin{equation}
\left.\frac{\partial G}{\partial r_+}\right|_{r_+=r_+^\star}=0;
\left.\frac{\partial^2 G}{\partial r_+^2}\right|_{r_+=r_+^\star}\gtrless 0
\ \ (\text{min/max}),
\label{eq:statability_general}
\end{equation}
so local stability is determined by the curvature of the landscape. 
Whenever a locally stable LBH minimum exists at $r_+=r_l>0$, we quantify (i) the global preference of the LBH relative to thermal AdS by the gap
\begin{equation}
\Delta G_{\rm glob}(T,\Lambda_e,\alpha)\equiv G(T,r_l)-G_{\rm AdS}=G(T,r_l),
\label{eq:DGglob_def}
\end{equation}
and (ii) the free-energy barrier separating the SBH saddle (local maximum) from the LBH minimum by
\begin{equation}
\Delta G_{\rm bar}(T,\Lambda_e,\alpha)\equiv G(T,r_s)-G(T,r_l),
\label{eq:DGbar_def}
\end{equation}
where $r_s$ denotes the SBH stationary point when it exists.
With these definitions, an \emph{overlap (coexistence) region} is characterized by
\begin{equation}
T\ge T_{\min}
\qquad \text{and} \qquad
\Delta G_{\rm glob}(T,\Lambda_e,\alpha)>0,
\label{eq:overlap_def}
\end{equation}
i.e., a locally stable LBH phase is available while thermal AdS remains globally preferred.\\
\indent The off-shell Gibbs free energy reads
\begin{equation}
G(T,r_+)=\frac{1}{6}r_+\Bigl[3-(1+\alpha)\Lambda_e r_+^2\Bigr]-\pi T\, r_+^2,
\label{eq:GI_offshell}
\end{equation}
and stationary points are determined by Eq.~\eqref{eq:statability_general}.
Equivalently, the stationarity condition can be written as $T=T_H^{\rm I}(r_+)$, where Hawking temperature is given by Eq.(\ref{TEMP1}).
For the ensemble temperature $T=T_{\rm H}(r_+)$ yields two branches 
\begin{align}
r_s(T)&=
\frac{-2\pi T+\sqrt{4\pi^2T^2+\Lambda_e}}{\sqrt{1+\alpha}\,\Lambda_e},\\
r_l(T)&=
\frac{-2\pi T-\sqrt{4\pi^2T^2+\Lambda_e}}{\sqrt{1+\alpha}\,\Lambda_e}.
\label{eq:rsrl_caseI}
\end{align}
respectively denoting SBH and LBH radii.  This happens only for a nonnegative discriminant,
\begin{equation}
4\pi^2T^2+\Lambda_e\ge 0
\quad \Longleftrightarrow \quad
T\ge T_{\min}\equiv \frac{\sqrt{-\Lambda_e}}{2\pi}.
\label{eq:TminI}
\end{equation}
Evaluating $\partial_{r_+}^2G$ at the stationary points confirms that $r_s$ is a local maximum (SBH saddle) and $r_l$ is a local minimum (LBH). 
 $\Delta G_{\rm glob}$ and $\Delta G_{\rm bar}$  are then given by 
\begin{multline}\label{DELTAGGLOB}
    \Delta G_{\rm glob}=-\frac{2\pi T+X}{2\sqrt{1+\alpha}\,\Lambda_e}
+\frac{(2\pi T+X)^3}{6\sqrt{1+\alpha}\,\Lambda_e^{2}}
\\ -\frac{\pi T\,(2\pi T+X)^2}{(1+\alpha)\,\Lambda_e^{2}}
\end{multline}
with $X \equiv \sqrt{4\pi^2 T^2+\Lambda_e}$ (see Appendix~\ref{app:DGbar_caseI}.), 
and (see Appendix \ref{GPMAP})
\begin{multline} \label{DELTAGBAR}
\Delta G_{\rm bar}(T,\Lambda_e,\alpha)
=\frac{2\,\sqrt{4\pi^2T^2+\Lambda_e}}{3\,\Lambda_e^{\,2}(1+\alpha)}
\big[
\sqrt{1+\alpha}\,\bigl(\Lambda_e-8\pi^2T^2\bigr)\\
+12\pi^2T^2\big].
\end{multline}

\paragraph*{(b) $T<T_{\min}$.}
In this case, the stationarity equation admits no real positive roots, so no black hole extrema exist in the landscape.
Indeed, $\partial_{r_+}G=0$ reduces to a quadratic whose discriminant is $4\pi^2(T)^2+\Lambda_e$, hence real roots require
$T\ge \sqrt{-\Lambda_e}/(2\pi)\equiv T_{\min}$ (at fixed $\Lambda_e$ this threshold is independent of $\alpha$).
Nevertheless, the off-shell functional $G(T,r_+)$ is still well-defined for trial $r_+>0$, and its explicit $\alpha$-dependence implies
$\left(\partial_\alpha G\right)_{T,r_+}=-(\Lambda_e/6)\,r_+^3>0$ for $\Lambda_e<0$, explaining the upward shift with increasing $\alpha$ in Fig.~\ref{Gmplot}(b), which becomes more pronounced at large $r_+$.

\paragraph*{(c) $T=T_{\min}$.}
At $T_{\min}$ the stationarity condition $\partial_{r_+}G=0$ has a vanishing discriminant, so the SBH saddle and LBH minimum merge into a single double root $r_s=r_l\equiv r_{\min}^{\rm I}$ (fold bifurcation) with $\partial_{r_+}G=\partial_{r_+}^2G=0$ and $\partial_{r_+}^3G\neq 0$, implying a locally flattened inflection-type stationary point rather than a regular minimum/maximum.
Consequently $G(T,r_s)=G(T,r_l)$ and the barrier height closes, $\Delta G_{\rm bar}(T_{\min})=0$ by definition; away from $r_{\min}^{\rm I}$ the large-$r_+$ behavior is controlled by the cubic term in $G(T,r_+)=\frac{r_+}{2}-\frac{(1+\alpha)\Lambda_e}{6}r_+^3-\pi T r_+^2$, so for $\Lambda_e<0$ one has $(\partial_\alpha G)_{T,r_+}=-(\Lambda_e/6)\,r_+^3>0$, explaining why the $\alpha>0$ curves bend upward and separate increasingly for larger $r_+$ in Fig.~6(c).

\begin{table*}
\caption{\label{tab:HP_compare}
Comparison of the Hawking--Page thermodynamic structure in Schwarzschild--AdS and Bumblebee--AdS gravity at fixed effective cosmological constant $\Lambda_e<0$.}
\begin{ruledtabular}
\begin{tabular}{ccc}
Quantity & Schwarzschild--AdS & Bumblebee--AdS \\
\hline
$T_{\min}$ 
& $\displaystyle \frac{\sqrt{-\Lambda}}{2\pi}$ 
& $\displaystyle \frac{\sqrt{-\Lambda_e}}{2\pi}$ \\

$T_{\rm HP}$ 
& $\displaystyle \frac{\sqrt{-\Lambda}}{\sqrt{3}\,\pi}$ 
& $\displaystyle \frac{\sqrt{(1+\alpha)(-\Lambda_e)}}{\sqrt{3}\,\pi}$ \\

$\mathcal{S}$ 
& $\displaystyle \frac{2}{\sqrt{3}}$ 
& $\displaystyle \frac{2}{\sqrt{3}}\sqrt{1+\alpha}$ \\

$\mathcal{W}$ 
& $\displaystyle \frac{2}{\sqrt{3}}-1$ 
& $\displaystyle \frac{2}{\sqrt{3}}\sqrt{1+\alpha}-1$ \\

Metastable window 
& standard 
& enlarged \\

$G_{\rm glob}$ 
& Schwarzschild limit 
& increases with $\alpha$ \\

$G_{\rm bar}$ 
& Schwarzschild limit 
& decreases with $\alpha$ \\
\end{tabular}
\end{ruledtabular}
\end{table*}

\paragraph*{(d) $T_{\min}<T<T_{\rm HP}$.}
In this window the stationarity equation $\partial_{r_+}G=0$ has two real positive roots $r_s$ and $r_l$ (discriminant $>0$), with $\partial_{r_+}^2G(r_s)<0$ and $\partial_{r_+}^2G(r_l)>0$, i.e. an SBH saddle and an LBH minimum, so $\Delta G_{\rm bar}>0$ and $\Delta G_{\rm glob}=G(T,r_l)>0$ (thermal AdS at $G_{\rm AdS}=0$ remains globally preferred).
The detailed \emph{shape} of the curves follows directly from the polynomial structure of $G$: the small-$r_+$ turnover is controlled by the competition between the linear and $-\,\pi T r_+^2$ terms, while the large-$r_+$ upturn is enforced by the AdS cubic term  whose explicit $\alpha$-variation, $(\partial_\alpha G)_{T,r_+}=-(\Lambda_e/6)\,r_+^3>0$, lifts the landscape increasingly with $r_+$, raising $G(T,r_l)$ more strongly than $G(T,r_s)$ and therefore increasing $\Delta G_{\rm glob}$ while reducing $\Delta G_{\rm bar}$ at fixed $(T,\Lambda_e)$.

\paragraph*{(e) $T=T_{\rm HP}$.}
The stationarity equation here retains two real roots,
\begin{equation}
r_s=\frac{1}{3}\,r_{\rm HP},
\qquad
r_l=r_{\rm HP},
\end{equation}
with $r_s$ the SBH saddle and $r_l$ the LBH minimum; correspondingly
$G(T_{\rm HP},r_l)=0$ while
$G(T_{\rm HP},r_s)=\frac{2}{9}r_s>0$,
so the barrier remains finite at coexistence,
$\Delta G_{\rm bar}(T_{\rm HP})=G(T_{\rm HP},r_s)$.
Crucially, since $r_{\rm HP}(\alpha)\propto (1+\alpha)^{-1/2}$, increasing $\alpha$
pushes the coexistence minimum to smaller $r_+$, exactly as seen in Fig.~6(e), while the
large-$r_+$ separation of the curves is governed by the explicit cubic LV term,
$(\partial_\alpha G)_{T,r_+}=-(\Lambda_e/6)\,r_+^3>0$.

\paragraph*{(f) $T>T_{\rm HP}$.}
For this case, LBH minimum is the globally preferred configuration, so $\Delta G_{\rm glob}=G(T,r_l)<0$.
The SBH saddle persists, hence the transition still involves a finite activation barrier $\Delta G_{\rm bar}=G(T,r_s)-G(T,r_l)>0$. 
The $\alpha$--trend is controlled by the LV deformation of the AdS (volume) term: at fixed $(T,r_+)$ one has
\begin{equation}
(\partial_\alpha G)_{T,r_+}=-\frac{\Lambda_e}{6}\,r_+^3>0\qquad(\Lambda_e<0)\,,
\end{equation}
so increasing $\alpha$ lifts the landscape more strongly at larger $r_+$. 
Along a stationary branch $r_\star\in\{r_s,r_l\}$, $\partial_{r_+}G(T,r_\star)=0$ implies
\begin{equation}
\frac{d}{d\alpha}G(T,r_\star)= (\partial_\alpha G)_{T,r_+}\Big|_{r_+=r_\star}
=-\frac{\Lambda_e}{6}\,r_\star^3\,,
\end{equation}
so both $G(T,r_s)$ and $G(T,r_l)$ increase with $\alpha$, with the LBH minimum lifted faster because $r_l>r_s$. 
Therefore $\Delta G_{\rm glob}$ increases (becomes less negative) and the barrier decreases,
\begin{equation}
\frac{d}{d\alpha}\Delta G_{\rm bar}
=-\frac{\Lambda_e}{6}\,(r_s^3-r_l^3)<0\,,
\end{equation}
which matches the progressive flattening of the barrier region and the relative upward shift of the large-$r_+$ part of the curves in Fig.~\ref{fig:Deltaglob}(f). \\
\indent In standard AdS black holes, the spacetime is asymptotically AdS, with the negative cosmological constant setting the infrared structure relevant for the Hawking--Page transition \cite{Dolan:2011xt}. Although the inclusion of Bumblebee \cite{Kostelecky:2003fs, Maluf:2020kgf} or Kalb--Ramond (KR) \cite{Yang:2023wtu, Liu:2024oas} fields does not alter the asymptotic AdS-like character of the geometry, as discussed in Sec.~\ref{geom}, the additional vector or tensor couplings modify the black-hole thermodynamics and hence deform the Hawking--Page landscape relative to the Schwarzschild--AdS case. The results obtained here indicate that the LV sector leaves a clear thermodynamic imprint even when the asymptotic AdS structure is preserved.

\indent  For convenience, we summarize the resulting on-shell and off-shell thermodynamic structure of the Schwarzschild--AdS and Bumblebee--AdS systems in Table~\ref{tab:HP_compare}. Table~\ref{tab:HP_compare} shows that the LV deformation leaves
$T_{\min}$ unchanged while shifting $T_{\rm HP}$ upward. Consequently, both
$\mathcal{S}$ and $\mathcal{W}$ increase with $\alpha$, implying a broader
interval in which the large-black-hole branch is locally stable but globally
disfavored relative to thermal AdS. The table makes explicit that the LV deformation leaves $T_{\min}$ unchanged while shifting $T_{\rm HP}$ and deforming the off-shell landscape, thereby enlarging the metastable Hawking--Page window.

\section{Summary and outlook}\label{summary}

\noindent We have analyzed the thermodynamic phase structure of neutral AdS black holes in Bumblebee gravity using complementary on-shell and off-shell methods. In the canonical ensemble at fixed effective pressure, the on-shell analysis shows that the Lorentz-violating (LV) parameter $\alpha$ shifts the Hawking--Page transition while preserving the standard small-/large-black-hole branching structure. A distinctive feature of the Bumblebee--AdS solution is that the Hawking--Page temperature $T_{\rm HP}$ increases monotonically with $\alpha$, whereas the minimum temperature $T_{\min}$ remains exactly independent of $\alpha$. As a consequence, the interval $T_{\min}<T<T_{\rm HP}$ is enlarged, implying a broader regime in which the large-black-hole branch is locally stable but globally disfavored relative to thermal AdS.\\
\indent The off-shell Gibbs free-energy landscape resolves this metastable structure explicitly. Treating the horizon radius as the order parameter, we identified the small-black-hole branch as a saddle and the large-black-hole branch as a local minimum whenever $T>T_{\min}$. This framework allows the metastable window to be quantified in terms of two analytic observables: the global-preference gap, which measures the free-energy difference between the large-black-hole minimum and thermal AdS, and the barrier height separating the saddle from the minimum. The resulting picture sharpens the standard Hawking--Page analysis by resolving not only the coexistence point but also the full LV deformation of the free-energy landscape.\\
\indent In the Lorentz-invariant limit $\alpha\to0$, all results reduce smoothly to the Schwarzschild--AdS case. The present analysis therefore identifies the free-energy landscape as a sensitive probe of LV deformations in asymptotically AdS black-hole thermodynamics.\\
\indent Several extensions are natural. It would be useful to generalize the analysis to charged or rotating LV AdS black holes, where richer phase structure is expected. A second direction is to study dynamical aspects of the metastable regime, including nucleation and decay rates associated with the off-shell barrier. These questions may help clarify whether LV sectors produce qualitatively new thermodynamic signatures beyond the neutral static case.

\appendix

\section{Cardano root for the horizon radius}\label{APPENDA}
The horizon radius $r_+$ is defined by $g_{tt}(r_+)=0$, i.e.
\begin{equation}
1-\frac{2M}{r_+}-\frac{(1+\alpha)\Lambda_e}{3}\,r_+^{2}=0
\Longleftrightarrow
(1+\alpha)\Lambda_e\,r_+^{3}-3r_+ + 6M=0 .
\end{equation}
A closed-form (Cardano) expression for the real root can be written as
\begin{equation}\label{eq:rplus_cardano_app}
r_{+}
=-\frac{\mathcal{F}(M,\Lambda_{e},\alpha)^{1/3}}{(1+\alpha)\Lambda_{e}}
-\frac{1}{\mathcal{F}(M,\Lambda_{e},\alpha)^{1/3}},
\end{equation}
where
\begin{equation}\label{eq:F_app}
\mathcal{F}(M,\Lambda_{e},\alpha)
\equiv
3(1+\alpha)^{2}\Lambda_{e}^{2}M
+\sqrt{(1+\alpha)^{3}\Lambda_{e}^{3}\Big(9(1+\alpha)\Lambda_{e}M^{2}-1\Big)} .
\end{equation}
Throughout we take the real branch of the cube root in \eqref{eq:rplus_cardano_app} and select the
largest positive root as the event-horizon radius for $\Lambda_e<0$.

\section{Global preference map} \label{GPMAP}
Define
\begin{equation}
X \equiv \sqrt{4\pi^2 T^2+\Lambda_e},
\qquad
r_{l}=\frac{-2\pi T- X}{\sqrt{1+\alpha}\,\Lambda_{e}} .
\end{equation}
We have 
\begin{align}
G(T,r_l^{\rm I})
&=\frac{1}{6} r_l\left[ 3 - (1+\alpha)\Lambda_e r_{l}^2 \right]
-\pi T r_l^2 \nonumber\\[2pt]
&=\frac{r_l}{2}-\frac{(1+\alpha)\Lambda_e}{6}r_l^3-\pi T r_l^2 \nonumber\\[4pt]
&= -\frac{2\pi T+X}{2\sqrt{1+\alpha}\,\Lambda_e}
+\frac{(2\pi T+X)^3}{6\sqrt{1+\alpha}\,\Lambda_e^{2}}
-\frac{\pi T\,(2\pi T+X)^2}{(1+\alpha)\,\Lambda_e^{2}} .
\end{align}

\section{Barrier height $\Delta G_{\rm bar}$}
\label{app:DGbar_caseI}
For $T>T_{\min}$,
\begin{align}
r_{\rm l,s}
&= \frac{T}{2\pi \sqrt{1+\alpha}\,\bigl(T_{\min}\bigr)^2}
\left[
1 \pm \sqrt{1 - \left(\frac{T_{\min}}{T}\right)^2}
\right], \label{eq:rlrs_given}\\[6pt]
S
&\equiv \sqrt{1 - \left(\frac{T_{\min}}{T}\right)^2}, \label{eq:S_def}\\
r_l
&=\frac{T}{2\pi \sqrt{1+\alpha}\,\bigl(T_{\min}\bigr)^2}(1+S),\\
r_s
&=\frac{T}{2\pi \sqrt{1+\alpha}\,\bigl(T_{\min}\bigr)^2}(1-S). \label{eq:rlrs_S}
\end{align}

The  off-shell Gibbs free energy is
\begin{align}\nonumber
G(T,r)
&=\frac{1}{6}\,r\Bigl[3-(1+\alpha)\Lambda_e r^2\Bigr]-\pi T r^2\\
&=\frac{r}{2}-\frac{(1+\alpha)\Lambda_e}{6}\,r^3-\pi T r^2 .
\label{eq:G_caseI}
\end{align}
The barrier height is defined by
\begin{equation}
\Delta G_{\rm bar}(T,\Lambda_e,\alpha)
=G\!\left(T,r_s\right)-G\!\left(T,r_l\right).
\label{eq:DGbar_def}
\end{equation}
Using Eq.~\eqref{eq:G_caseI},
\begin{equation}
\Delta G_{\rm bar}
=\frac{1}{2}(r_s-r_l)
-\frac{(1+\alpha)\Lambda_e}{6}(r_s^3-r_l^3)
-\pi T (r_s^2-r_l^2),
\label{eq:DGbar_expand}
\end{equation}
where $r_{l,s}\equiv r_{l,s}$.

\begin{align}\nonumber
r_s-r_l
&=\frac{T}{2\pi \sqrt{1+\alpha}\,\bigl(T_{\min}\bigr)^2}\Bigl[(1-S)-(1+S)\Bigr]\\
&=-\frac{T}{\pi \sqrt{1+\alpha}\,\bigl(T_{\min}\bigr)^2}\,S, \label{eq:rs_minus_rl}\\[6pt]
\nonumber r_s+r_l
&=\frac{T}{2\pi \sqrt{1+\alpha}\,\bigl(T_{\min}\bigr)^2}\Bigl[(1-S)+(1+S)\Bigr]\\
&=\frac{T}{\pi \sqrt{1+\alpha}\,\bigl(T_{\min}\bigr)^2}, \label{eq:rs_plus_rl}\\[6pt]
r_s^2-r_l^2
\nonumber &=(r_s-r_l)(r_s+r_l)\\
&=-\frac{T^2}{\pi^2(1+\alpha)\,\bigl(T_{\min}\bigr)^4}\,S. \label{eq:rs2_minus_rl2}
\end{align}
Next, using $r_s^3-r_l^3=(r_s-r_l)\bigl(r_s^2+r_s r_l+r_l^2\bigr)$, we compute:

\begin{align}
r_s r_l
&=
\left(\frac{T}{2\pi \sqrt{1+\alpha}\,\bigl(T_{\min}\bigr)^2}\right)^2(1-S)(1+S)\\
&=
\left(\frac{T}{2\pi \sqrt{1+\alpha}\,\bigl(T_{\min}\bigr)^2}\right)^2(1-S^2) \nonumber\\
&=
\left(\frac{T}{2\pi \sqrt{1+\alpha}\,\bigl(T_{\min}\bigr)^2}\right)^2
\left(\frac{\bigl(T_{\min}\bigr)^2}{T^2}\right)\\
&=
\frac{1}{4\pi^2(1+\alpha)\,\bigl(T_{\min}\bigr)^2}, \label{eq:rsrl}
\end{align}
\begin{align}
r_s^2+r_l^2
&=(r_s+r_l)^2-2r_s r_l\\
&=\left(\frac{T}{\pi \sqrt{1+\alpha}\,\bigl(T_{\min}\bigr)^2}\right)^2
-2\left[\frac{1}{4\pi^2(1+\alpha)\,\bigl(T_{\min}\bigr)^2}\right] \nonumber\\
&=
\frac{1}{\pi^2(1+\alpha)\,\bigl(T_{\min}\bigr)^4}
\left(T^2-\frac{\bigl(T_{\min}\bigr)^2}{2}\right), \label{eq:rs2_plus_rl2}\\[6pt]
r_s^2+r_s r_l+r_l^2
\nonumber &=
\frac{1}{\pi^2(1+\alpha)\,\bigl(T_{\min}\bigr)^4}
\left(T^2-\frac{\bigl(T_{\min}\bigr)^2}{2}\right)\\
\nonumber &+\frac{1}{4\pi^2(1+\alpha)\,\bigl(T_{\min}\bigr)^2} \nonumber\\
&=
\frac{1}{\pi^2(1+\alpha)\,\bigl(T_{\min}\bigr)^4}
\left(T^2-\frac{\bigl(T_{\min}\bigr)^2}{4}\right), \label{eq:rs2_plus_rsrL_plus_rl2}
\end{align}
\begin{align}
r_s^3-r_l^3
&=(r_s-r_l)\bigl(r_s^2+r_s r_l+r_l^2\bigr) \nonumber\\
\nonumber &=
-\frac{T}{\pi \sqrt{1+\alpha}\,\bigl(T_{\min}\bigr)^2}S 
\frac{1}{\pi^2(1+\alpha)\,\bigl(T_{\min}\bigr)^4}\\
& \qquad \times \left(T^2-\frac{\bigl(T_{\min}\bigr)^2}{4}\right) \nonumber\\
&=
-\frac{T\left(T^2-\frac{\bigl(T_{\min}\bigr)^2}{4}\right)}
{\pi^3(1+\alpha)^{3/2}\,\bigl(T_{\min}\bigr)^6}\,S. \label{eq:rs3_minus_rl3}
\end{align}

Substituting Eqs.~\eqref{eq:rs_minus_rl}, \eqref{eq:rs2_minus_rl2}, and \eqref{eq:rs3_minus_rl3} into Eq.~\eqref{eq:DGbar_expand}, we obtain
\begin{widetext}
\begin{align}
\Delta G_{\rm bar}(T,\Lambda_e,\alpha)
&=
-\frac{T}{2\pi \sqrt{1+\alpha}\,\bigl(T_{\min}\bigr)^2}\,S
+\frac{\Lambda_e\,T\left(T^2-\frac{\bigl(T_{\min}\bigr)^2}{4}\right)}
{6\pi^3\sqrt{1+\alpha}\,\bigl(T_{\min}\bigr)^6}\,S
+\frac{T^3}{\pi(1+\alpha)\,\bigl(T_{\min}\bigr)^4}\,S \nonumber\\[4pt]
&=
S\left[
-\frac{T}{2\pi \sqrt{1+\alpha}\,\bigl(T_{\min}\bigr)^2}
+\frac{\Lambda_e\,T\left(T^2-\frac{\bigl(T_{\min}\bigr)^2}{4}\right)}
{6\pi^3\sqrt{1+\alpha}\,\bigl(T_{\min}\bigr)^6}
+\frac{T^3}{\pi(1+\alpha)\,\bigl(T_{\min}\bigr)^4}
\right].
\label{eq:DGbar_Tmin_form}
\end{align}
Now using 
\begin{equation}
S=\sqrt{1-\frac{\bigl(T_{\min}\bigr)^2}{T^2}}
=\sqrt{1+\frac{\Lambda_e}{4\pi^2T^2}}
=\frac{\sqrt{4\pi^2T^2+\Lambda_e}}{2\pi T}.
\label{eq:S_Lambdae}
\end{equation}
Using  (\ref{eq:S_Lambdae}) in the above expression, and simplifying yields Eq.~(\ref{DELTAGBAR}).
\end{widetext}

\bibliographystyle{apsrev4-1}
\bibliography{LIVBH.bib}
\end{document}